\documentclass{article} % For LaTeX2e
\usepackage{iclr2027_conference,times}

\usepackage{amsmath,amsfonts,bm}

\def\eqref#1{equation~\ref{#1}}
\def\1{\bm{1}}

\DeclareMathAlphabet{\mathsfit}{\encodingdefault}{\sfdefault}{m}{sl}
\SetMathAlphabet{\mathsfit}{bold}{\encodingdefault}{\sfdefault}{bx}{n}

\usepackage{hyperref}
\usepackage{url}
\usepackage{graphicx}
\usepackage{subcaption}

\title{LiftDiff: Guided Discrete Diffusion for Quantum Code Construction}

\author{Yotam Peled, Eliya Nachmani \\
School of Electrical And Computer Engineering\\
Ben-Gurion University of the Negev, Israel\\
\texttt{peledyot@post.bgu.ac.il, eliyanac@bgu.ac.il} \\
}

\iclrfinalcopy % Uncomment for camera-ready version, but NOT for submission.
\begin{document}

\maketitle

\begin{abstract}
The realization of fault-tolerant quantum computation relies on Quantum
Error-Correcting Codes (QECC). Of particular interest are codes that combine a high encoding rate with sufficient distance. Existing algebraic constructions, such as lifted product (LP) codes, build a large quantum low-density parity-check (qLDPC) code from a classical base matrix. In this work we introduce LiftDiff, a generative approach to the discovery of QECC. A discrete denoising diffusion model is trained to generate base matrices over the group algebra
$\mathbb{F}_2[\mathbb{Z}_\ell]$, which are then expanded into LP codes. The reverse diffusion process is guided by an energy function that evaluates quantum code properties, including girth, degree, sparsity and decoding performance, steering the sampler toward matrices that yield LP codes with favorable parameters. Our method discovers codes with parameters such as $[[169,13,11]]$, $[[143,15,9]]$, and $[[260,24,10]]$, including a $[[247,19,14]]$ code that achieves the highest ${kd^2}/{n}$ among the baselines. Moreover, several discovered codes outperform the Gross code across a broad range of the tested physical error rates under both code-capacity and circuit-level noise.

\end{abstract}

\section{Introduction}
Quantum computing harnesses the laws of quantum mechanics to perform computation, giving rise to a fundamentally different computational paradigm from classical computing. This paradigm has the potential to provide substantial computational advantages for problems that are difficult to solve using classical methods
\citep{Shor_1997, grover1996fast, gidney2021factor, feynman2018simulating}. Quantum systems are, however, exceptionally noisy. The qubit, the elementary unit of quantum information, is coupled to its environment, and the resulting decoherence destroys the superpositions and phase relations on which the computation depends. Errors therefore accumulate before a computation of useful length can be completed. Quantum Error Correction (QEC) \citep{PhysRevA.54.1098} is thus indispensable for the realization of a quantum computer. Shor's first QEC scheme \citep{PhysRevA.52.R2493} demonstrated that these obstacles can be overcome, and that quantum information can be protected despite the impossibility of copying it and the continuous nature of the errors that afflict it. Considerable progress has been made since through the stabilizer formalism \citep{gottesman1997stabilizer}. Within the stabilizer framework, Calderbank-Shor-Steane (CSS) codes \citep{steane1996error, PhysRevA.54.1098} form an important class of QECC constructed from two parity check matrices satisfying an orthogonality condition. Their structure separates the correction of bit-flip and phase-flip errors. Among the QEC schemes proposed, the surface code \citep{kitaev1997quantum, bravyi1998quantumcodeslatticeboundary, Kitaev_2003, Dennis_2002} has received the most attention. It tolerates a comparatively high physical error rate, and efficient decoders for it are available. Its drawback is the encoding rate. A surface code patch encodes a single logical qubit into a number of physical qubits that grows quadratically with the distance, so its encoding rate vanishes as the protection is increased.
This has motivated the search for codes that keep the parity checks sparse while encoding many logical qubits in the same block. Classical LDPC codes combine sparse checks with efficient message-passing decoding
and performance close to capacity, and recent years have seen sustained effort to create qLDPC codes  \citep{Breuckmann_2021}, where the checks must in addition commute. The Hypergraph Product (HGP) \citep{Tillich_2014} provides a way to construct a qLDPC code using the parity check matrix of two classical codes. The LP construction replaces the binary entries of the base matrix with elements of a group algebra \citep{Panteleev_2022}, allowing the constructed codes to achieve high encoding rate and distance. Among the most promising qLDPC constructions are the Bivariate Bicycle (BB) \citep{Bravyi_2024} codes, whose CSS structure is defined through a pair of polynomials in the ring $\mathbb{F}_2[x,y]/(x^{\ell}-1,y^{m}-1)$. Machine learning has been applied to a range of problems in the domain of quantum computing, such as quantum error decoding \citep{wang2024artificialintelligencequantumerror, bausch2024learning, senior2026scalablerealtimeneuraldecoder,chamberland2026fast} and the design of QECC \citep{olle2024simultaneous, PhysRevApplied.23.034048, liu2026largelanguagemodeldiscoveryquantumldpc, yan2026omniqecdiscoveringpracticalquantum,cruzbenito2026evolutionarydiscoverybivariatebicycle}. Several approaches have been used to search the space of QECC. However, the discrete and combinatorial nature of the design space, together with the costly evaluation of the resulting codes, makes efficiently navigating this space challenging, particularly for larger constructions. Generative models offer another approach to this problem. A generative model can learn the underlying structure of the design space and produce new candidate constructions.

In this work, we introduce LiftDiff, a generative approach to the discovery of QECC. Our method employs a discrete diffusion model trained on classical base matrices, which are subsequently used to construct quantum codes through LP. During inference, we guide the diffusion process using a set of energy functions designed to promote desirable code properties, steering the generation toward promising quantum code constructions. Our contributions are as follows: 
\begin{itemize}
    \item We introduce, to the best of our knowledge, the first discrete diffusion framework for the generative construction of novel QECC.
    
    \item We develop an energy-based guidance mechanism that operates throughout the reverse diffusion process, using differentiable quantum code properties to steer the generation toward base matrices with desirable properties after lifting. 
    
    \item LiftDiff can generate QECC with diverse combinations of code parameters, including $[[143,15,9]]$, $[[260, 24, 10]]$, and $[[169,13,11]]$, as well as a $[[247,19,14]]$ code that achieves the highest $kd^2/n$ among all baselines for a check weight $\leq 8$.
\end{itemize}

\section{Related Work}
Two major topics of research in QEC \citep{Roffe_2019} are construction of quantum codes and the development of algorithms for decoding them, with recent work increasingly exploring machine learning as a tool for both tasks \citep{wang2024artificialintelligencequantumerror, Alexeev_2025, Krenn_2023, yan2026rethinkroleneuraldecoders}. The construction of quantum codes has traditionally relied on algebraic and combinatorial methods, leading to a variety of code families with different trade-offs between rate, distance, and structural properties \citep{Bravyi_2024, Tillich_2014, Panteleev_2022, Breuckmann_2021, Kasai_2026, zhao2026ultrahighratequantumerrorcorrection, Breuckmann_2016}. In terms of decoding, a wide range of decoders have been developed, such as Minimum Weight Perfect Matching, Union Find and Belief Propagation with Ordered Statistics Decoding (BP-OSD) \citep{edmonds1965paths, fowler2014minimumweightperfectmatching, Delfosse_2021, Panteleev_2021}, as well as their variants \citep{Wu_2025, peled2026neuralminimumweightperfect, kwak2026logicalneuralbeliefpropagation}. More recently machine learning was also applied to develop neural decoders, spanning from Convolutional Neural Networks (CNN) \citep{gu2026scalableneuraldecoderspractical}, Graph Neural Networks (GNN) \citep{Lange_2025} to Transformers \citep{Blue_2026, zenati2026saqstabilizerawarequantumerror,choukroun2023deepquantumerrorcorrection}. Code Design was also explored using machine learning techniques, starting from reinforcement learning \citep{olle2024simultaneous, Nautrup_2019, PhysRevApplied.23.034048, sella2026designingquantumerrorcorrecting}, bayesian optimization \citep{chengyu2026bayesianoptimizationquantumerrorcorrecting} to the usage of Large Language Models (LLMs) \citep{cruzbenito2026evolutionarydiscoverybivariatebicycle, yan2026omniqecdiscoveringpracticalquantum}.

\section{Background}

\textbf{Quantum Foundations.} A qubit is the basic unit of quantum information and can be described by a state: $|\psi\rangle = \alpha |0\rangle + \beta |1\rangle$ where $\alpha,\beta \in \mathbb{C}$ are probability amplitudes satisfying
$|\alpha|^2+|\beta|^2=1$. An $n$-qubit system is described by a state in a
$2^n$-dimensional Hilbert space. QECC exploit this
structure to encode logical information into a larger set of physical qubits
and protect it from noise. \\ \textbf{Stabilizer Formalism.} The stabilizer formalism \citep{gottesman1997stabilizer} encodes quantum information by constraining an $n$-qubit system to the simultaneous $+1$ eigenspace of a commuting set of Pauli operators within the Hilbert space $\mathcal{H}_2^n$. These operators are drawn from the $n$-qubit Pauli group $\mathcal{P}_n$, which consists of tensor products of single-qubit Pauli matrices. Errors manifest as operators that violate these symmetry constraints and can be identified through projective measurements that do not disturb the encoded logical information. Concretely, an $[[n,k,d]]$ quantum code encodes $k$ logical qubits into $n$ physical qubits with distance $d$, the stabilizer code is specified by choosing $m = n - k$ independent generators $\{S_i\}_{i=1}^m$ subject to the constraint that the group generated by these operators does not contain $-I$. The associated logical code space is given by $C_S = \left\{\, |\psi\rangle \in \mathcal{H}_2^n \;\middle|\; S_i |\psi\rangle = |\psi\rangle,\ \forall i \in \{1,\dots,m\} \right\}$. Measuring the stabilizer generators produces a binary syndrome that reflects the commutation relations between an error operator and the stabilizers, thereby enabling error identification while preserving the logical state, and a set of commuting stabilizer generators can be represented by a binary check matrix.\\
\textbf{CSS Codes.} CSS codes are a subclass of stabilizer codes in which the stabilizer generators are decoupled into $X$-type and $Z$-type checks. They can therefore be described by two binary matrices $H_X$ and $H_Z$, which must satisfy $H_X H_Z^T = 0$
to ensure that all stabilizer generators commute. In this case, the number of encoded logical qubits is: $k = n-\text{rank}(H_X) - \text{rank}(H_Z)$ \\
\textbf{Lifted Product Codes.} The LP construction \citep{Panteleev_2022} provides a systematic way of constructing
CSS quantum codes from matrices over a group algebra. Let
\begin{equation}
    M_1 \in R^{m_1 \times n_1},
    \qquad
    M_2 \in R^{m_2 \times n_2},
\end{equation}
where $R$ is a group algebra over $\mathbb{F}_2$. The construction combines
$M_1$ and $M_2$ and their involutions to form the CSS check matrices
\begin{equation}
    H_X =
    \left[
    M_1 \otimes I_{n_2}
    \;\middle|\;
    I_{m_1} \otimes M_2^{\ast}
    \right],
    \qquad
    H_Z =
    \left[
    I_{n_1} \otimes M_2
    \;\middle|\;
    M_1^{\ast} \otimes I_{m_2}
    \right],
\end{equation}
where $M^\ast$ denotes the transpose together with the involution on the
group algebra. By construction, the resulting matrices satisfy
$H_XH_Z^{\ast}=0$ and therefore define a CSS code. For the cyclic group algebra
$R=\mathbb{F}_2[x]/(x^\ell-1)$, each entry of the base matrices is a
polynomial of degree less than $\ell$, and the lifted representation replaces
each polynomial by its corresponding $\ell\times\ell$ circulant binary
matrix. In this work, we consider the self-product case
$LP(M,M)$.

\section{Method}
\label{sec: method}
\subsection{Overview}
We propose a novel framework for constructing QECCs, specifically LP codes. LiftDiff does not operate directly on parity-check matrices, instead, it operates on a base matrix $M \in R^{m \times n}$ over
the group algebra $R = \mathbb{F}_2[\mathbb{Z}_\ell]$. The matrix $M$ then defines a quantum code via $LP(M,M)$. The QECC search space is enormous, making the search for codes with specific properties challenging. Discrete diffusion is particularly well suited to this task for two reasons: (i) codes are binary, making discrete diffusion over binary instances a natural choice. (ii) The iterative nature of the diffusion process, together with its differentiable formulation, allows us to specify desired code properties and guide the diffusion process toward high-quality codes.  Our pipeline consists of three stages:

\textbf{(i) Training.} A discrete denoising diffusion model is trained to invert
a bit-flip corruption process on base matrices drawn from a
progressive-edge-growth (PEG) construction. The model learns what a sparse, high-girth base matrix looks like.

\begin{figure}[t!]
  \centering
  \includegraphics[width=0.95\linewidth, height=3.3cm]{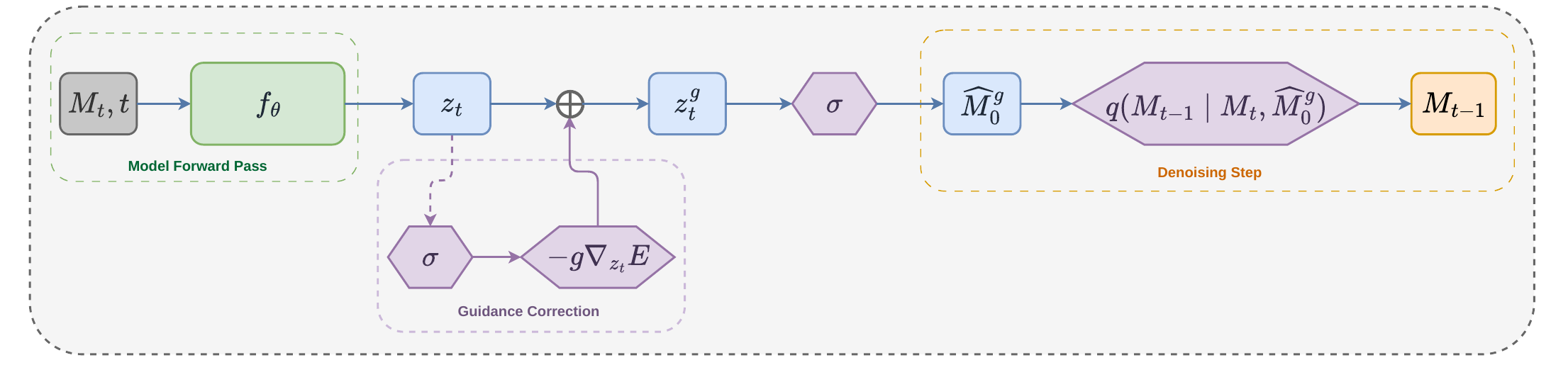}
  \caption{Overview of the inference process. The matrix $M_t$ and timestep $t$ are input to the model, producing logits $z_t$. The logits are passed through a sigmoid function and fed into the guidance mechanism, whose output adjusts the prediction to obtain $z_t^{g}$. The guided logits are passed through a second sigmoid function, and the resulting output is used to produce $M_{t-1}$.
}
  \label{fig: workflow}
\end{figure}

\textbf{(ii) Guided inference.} The trained model is sampled in reverse, and at
every denoising step the prediction is displaced along the gradient of an energy that scores the quantum code the current base matrix implies. The energy is
computed by pushing the model's continuous prediction through the LP. This stage produces batches of base matrices. This process is illustrated in Figure \ref{fig: workflow}

\textbf{(iii) Screening and certification.} Every sample is expanded, verified to
commute, and given its exact $n$ and $k$, a relatively cheap distance bound \citep{Pryadko_2022} is used to give a first indication of each code. The distance of the best codes is finally proven using Mixed-Integer Linear Programming (MILP) \citep{cruzbenito2026evolutionarydiscoverybivariatebicycle}.

\subsection{The generative model}
\label{sec:model}

\paragraph{Corruption process.}
Let $M_0 \in \{0,1\}^{m \times n \times \ell}$ be a clean base matrix, written as
a tensor whose last axis holds the $\ell$ coefficients of each cell's
polynomial. The forward process is the discrete diffusion of D3PM \citep{austin2021structured} with a symmetric binary channel: at step $t$ every
bit is flipped independently with probability $\beta_t$. Composing $t$ steps
gives a single Bernoulli channel,
\begin{equation}
    M_t = M_0 \oplus \varepsilon_t,
    \qquad
    \varepsilon_t \sim \mathrm{Bernoulli}(1 - \bar\alpha_t)^{\otimes mn\ell},
    \qquad
    \bar\alpha_t = \tfrac12\Bigl(1 + \prod_{s \le t}(1 - 2\beta_s)\Bigr),
\end{equation}
where $\bar\alpha_t$ is the probability that a given bit has survived unflipped.
As $t \to T$, $\bar\alpha_t \to \tfrac12$ and $M_T$ is uniform noise. We use
either a linear schedule on $\beta_t$ or a cosine schedule on $\bar\alpha_t$
rescaled to $(\tfrac12, 1]$, which is the reachable range for a binary symmetric
channel.

\subsection{Model Architecture}

The backbone of our model is a message passing architecture utilizing cross-attention mechanisms over a bipartite graph \citep{vaswani2017attention}. Let the base matrix be a tensor $M \in \mathbb{F}_2^{m \times n \times \ell}$, where each of the $\ell$ entries in a cell indicates whether the corresponding term appears in the polynomial. Given a noisy tensor $M_t$ at diffusion timestep $t$, the model aims to predict the clean tensor $M_0$. The architecture initializes learnable node embeddings for the check and variable nodes, $Z_C^{(0)} \in \mathbb{R}^{m \times d}$ and $Z_V^{(0)} \in \mathbb{R}^{n \times d}$, where $d$ is the hidden dimension. The scalar timestep $t$ is mapped to a continuous time embedding $c \in \mathbb{R}^d$ via sinusoidal positional encodings \citep{vaswani2017attention} followed by a multi-layer perceptron (MLP). To incorporate the structural information of the noisy graph, $M_t$ is collapsed along its $\ell$-dimension to form a bipartite adjacency matrix $\tilde{M}_t \in \mathbb{R}^{m \times n}$. For each check node $i$ and variable node $j$, the $\ell$-dimensional vector $(M_t)_{i,j}$ is projected via a learned linear layer:
\begin{equation}
    \tilde{M}_{t,i,j} = \mathbf{w}^T (M_t)_{i,j} + b,
\end{equation}
where $\mathbf{w} \in \mathbb{R}^\ell$ and $b \in \mathbb{R}$. The core of the network consists of $L$ bipartite attention blocks. In each block, node representations are first modulated by the time embedding $c$ using Adaptive Layer Normalization (AdaLN). Specifically, $c$ is projected via a weight matrix $W_{\text{1}} \in \mathbb{R}^{d \times 4d}$ to generate scale and shift parameters $\gamma_C, \beta_C, \gamma_V, \beta_V \in \mathbb{R}^d$:
\begin{equation}
    [\gamma_C, \beta_C, \gamma_V, \beta_V] = W_{\text{1}}(\operatorname{SiLU}(c)),
\end{equation}
which are used to normalize the check and variable node representations:
\begin{equation}
    \hat{Z}_C = \operatorname{LayerNorm}(Z_C) \odot (1 + \gamma_C) + \beta_C, \quad
    \hat{Z}_V = \operatorname{LayerNorm}(Z_V) \odot (1 + \gamma_V) + \beta_V.
\end{equation}
During the variable-to-check ($V \to C$) message passing phase, check nodes attend to variable nodes. To account for multi-head attention with $H$ heads, let $d_h = d/H$ be the head dimension. For each head $h \in \{1,\dots,H\}$, the queries, keys, and values are explicitly computed from the normalized representations: $Q_C^{(h)} = \hat{Z}_C W_{Q,h}^C, K_V^{(h)} = \hat{Z}_V W_{K,h}^V, V_V^{(h)} = \hat{Z}_V W_{V,h}^V$
where $W_{Q,h}^C, W_{K,h}^V, W_{V,h}^V \in \mathbb{R}^{d \times d_h}$ are the weight matrices for head $h$, resulting in $Q_C^{(h)} \in \mathbb{R}^{m \times d_h}$ and $K_V^{(h)}, V_V^{(h)} \in \mathbb{R}^{n \times d_h}$. The structural bias is injected into the pre-softmax affinities of each head:
\begin{equation}
    U_C^{(h)} = \operatorname{Softmax}\left( \frac{Q_C^{(h)} (K_V^{(h)})^T}{\sqrt{d_h}} + \alpha^{(h)} \tilde{M}_t \right) V_V^{(h)},
\end{equation}
where $\alpha^{(h)} > 0$ is a time-dependent learned scalar specific to head $h$. The outputs $U_C^{(h)}$ from all $H$ heads are concatenated to form the combined update $U_C \in \mathbb{R}^{m \times d}$. The node representations are then updated via an MLP and a residual connection:
\begin{equation}
    Z_C^{\text{out}} = Z_C + \operatorname{MLP}_C(U_C).
\end{equation}
A symmetric operation updates the variable nodes ($C \to V$) using $\tilde{M}_t^T$ to yield $Z_V^{\text{out}}$. After $L$ iterations and a final LayerNorm, the representations $Z_C^{(L)}$ and $Z_V^{(L)}$ are processed by an edge decoder. We employ an independent bilinear map for each polynomial coefficient index $k \in \{1,\dots,\ell\}$:
\begin{equation}
    \text{Logits}_{i,j,k} = \frac{1}{\sqrt{d}} \left(Z_{C,i}^{(L)}\right)^T W^{(k)} Z_{V,j}^{(L)}, \quad W^{(k)} \in \mathbb{R}^{d \times d}.
\end{equation}
Finally, a sigmoid activation $\sigma(\cdot)$ is applied to output the predicted tensor $\hat{M}_0 \in (0,1)^{m \times n \times \ell}$:
\begin{equation}
    \widehat{M}_0 = (\hat{M}_0)_{i,j,k} = \sigma(\text{Logits}_{i,j,k})
\end{equation}

\paragraph{Training objective.}
Training is the standard $x_0$-prediction objective, we draw $M_0$ from the data,
draw $t$ uniformly, corrupt, and ask the network to recover $M_0$. Since every
entry is a bit, the reconstruction term is a binary cross-entropy over all
$mn\ell$ coefficients. To this we add a degree term $d_v$ to gain control over the final structure. The degree term is computed on the \emph{support}, not on the coefficients, we collapse the polynomial axis by using a differentiable soft-OR, the probability that cell $(i,j)$ is nonzero: $\widehat{S}_{ij} = 1 - \prod_{k}\bigl(1 - \widehat{M}_{0,ijk}\bigr)$.
Thus, the final training objective is:
\begin{equation}
    \mathcal{L} =
    \mathrm{BCE}\bigl(\widehat{M}_0,\, M_0\bigr)
    + \lambda_{\deg}\,
    \Bigl(
    \bigl\| \mathrm{col}(\widehat{S}) - d_v \bigr\|^2_{\text{mean}}
    + \bigl\| \mathrm{row}(\widehat{S}) - \tfrac{n d_v}{m} \bigr\|^2_{\text{mean}}
    \Bigr)
    \label{eq:train-loss}
\end{equation}

\subsection{Guided inference}
\label{sec:guidance}
To generate new codes, and to optimize their properties, we guide the diffusion process along its whole trajectory. We define an energy $E$ as a weighted sum of sub-energies, $E \;=\; \sum_i \lambda_i E_i$
where each $E_i$ represents one property of the code that we wish to optimize and $\lambda_i \ge 0$ sets its strength.
Unguided sampling draws the next state from the D3PM posterior using
the network prediction. Writing $z_t = f_\theta(M_t, t)$ for the logits the network produces at step $t$, one reverse step is
\begin{equation}
    z_t = f_\theta(M_t, t), \qquad
    \widehat{M}_0 = \mathrm{sigmoid}(z_t), \qquad
    M_{t-1} \sim q\bigl(M_{t-1} \mid M_t,\, \widehat{M}_0\bigr).
    \label{eq:unguided}
\end{equation}
We modify \eqref{eq:unguided} to
\begin{equation}
    \begin{aligned}
        z_t &= f_\theta(M_t, t), & \quad
        z_t^{g} &= z_t - g\,\nabla_{z_t} E\bigl(\mathrm{sigmoid}(z_t)\bigr), \\
        \widehat{M}_0^{g} &= \mathrm{sigmoid}\bigl(z_t^{g}\bigr), & \quad
        M_{t-1} &\sim q\bigl(M_{t-1} \mid M_t,\, \widehat{M}_0^{g}\bigr).
    \end{aligned}
    \label{eq:guided}
\end{equation}
where $g$ is a tunable hyperparameter defined as the guidance scale. The guidance is performed in the logit space rather
than probabilities. Logits are unconstrained unlike probabilities. Nothing about the corruption process is modified. The forward process, the
schedule and the posterior $q$ are exactly those of Section~\ref{sec:model}, the guidance enters only through $\widehat{M}_0$ which the posterior is conditioned on.

\paragraph{Energy Terms}
The energy scores the code implied by a continuous base matrix. Two objects are used, and each term below is defined on one of them explicitly. Pushing
$\widehat{M}_0 \in [0,1]^{m\times n\times\ell}$ through the LP with the mod-2 reduction dropped gives real-valued check matrices $H_X, H_Z$, applying
the lift alone gives the \emph{unfolded base matrix}
\begin{equation}
    \mathcal{M} \;=\; {unfold}\bigl(\widehat{M}_0\bigr) \;\in\; \mathbb{R}^{\bar m \times \bar n},
    \qquad \bar m = m\ell, \quad \bar n = n\ell,
\end{equation}
the Tanner matrix of the classical base graph, whose rows are check nodes and
whose columns are variable nodes. Parities of real numbers are taken with
\begin{equation}
    \mathrm{par}(s) \;=\; \tfrac12\bigl(1 - \cos(\pi s)\bigr),
    \label{eq:par}
\end{equation}
which returns $s \bmod 2$ whenever $s$ is an integer and interpolates smoothly in
between, so a syndrome can be formed from real-valued matrices without breaking
the gradient.

\paragraph{1. Decoding}
The first term asks that the generated code be easy to decode. We simulate the procedure of the decoding task. We sample a depolarizing error at physical rate $p$ under the code-capacity noise model. Writing $e_X, e_Z \in \{0,1\}^{n}$ for the two components of that error, the syndromes are formed against the continuous
check matrices through \eqref{eq:par},
\begin{equation}
    s_Z = \mathrm{par}\bigl(e_X H_Z^{\mathsf{T}}\bigr),
    \qquad
    s_X = \mathrm{par}\bigl(e_Z H_X^{\mathsf{T}}\bigr).
\end{equation}
A BP decoder is then unrolled \citep{nachmani2016learning} for a fixed number of iterations,
with damping and a constant channel log-likelihood ratio
$\lambda_{\mathrm{ch}} = \log\frac{1 - 2p/3}{2p/3}$, producing soft estimates
$\widehat{e}_X = \mathrm{BP}(H_Z, s_Z)$ and
$\widehat{e}_Z = \mathrm{BP}(H_X, s_X)$. Unrolling is what keeps the decoder
differentiable, every message update is an arithmetic operation on $H_X$ and
$H_Z$, so the error the decoder makes is backpropagated through the decoding
dynamics, through the LP, and into the model's logits. The term is
the decoder's residual uncertainty about the true error (see Appendix \ref{app:bp}):
\begin{equation}
    E_{\mathrm{LER}} \;=\;
    \tfrac12\Bigl[
      \mathrm{BCE}\bigl(\widehat{e}_X,\, e_X\bigr)
    + \mathrm{BCE}\bigl(\widehat{e}_Z,\, e_Z\bigr)
    \Bigr],
\end{equation}
\paragraph{2. Geometry} The next four terms act on the unfolded base matrix $\mathcal{M}$. We evaluated them on $H_X$ and $H_Z$ as well and no difference was found in the resulting codes, therefore we used $\mathcal{M}$ for more efficient compute. Let us define the following matrices:
\begin{equation}
    A_c = \mathcal{M}\mathcal{M}^{\mathsf{T}} \in \mathbb{R}^{\bar m \times \bar m},
    \qquad
    A_v = \mathcal{M}^{\mathsf{T}}\mathcal{M} \in \mathbb{R}^{\bar n \times \bar n}
\end{equation}
for the two node-projection matrices, an off-diagonal entry of $A_c$ counts the variable nodes shared by a pair of checks, and symmetrically for $A_v$. Short cycles can affect BP performance, so the second and third terms suppress them. A $4$-cycle is a pair of nodes sharing two neighbours, which is
exactly an off-diagonal entry of $A_c$ or $A_v$ exceeding one, so
\begin{equation}
    E_{\mathrm{girth}_4} \;=\;
    \frac{\sum_{i \neq j} (A_c)_{ij}^2 \;+\; \sum_{i \neq j} (A_v)_{ij}^2}
         {\bar m(\bar m - 1) + \bar n(\bar n - 1)} ,
\end{equation}
normalized by the number of ordered off-diagonal pairs for a size-independent scale. \\
The third term suppresses $6$-cycles. Writing
$B_c = A_c \odot (\mathbf{1} - I)$ and $B_v = A_v \odot (\mathbf{1} - I)$ for the
projections with diagonals removed, $\operatorname{tr}(B^3)$ counts closed
triples, but over-counts. A single node of degree three or more creates triangles that are not bipartite $6$-cycles. We therefore subtract those spurious contributions. We define: $S_r^v = \sum_i \mathcal{M}_{ij}^{2r}$ per column and
$S_r^c$ the analogous row quantity,
\begin{equation}
    E_{\mathrm{girth}_6} =
    \frac{\bigl[\operatorname{tr}(B_c^3) - \Sigma_c\bigr]_+}{\bar m(\bar m -1)(\bar m -2)}
    +
    \frac{\bigl[\operatorname{tr}(B_v^3) - \Sigma_v\bigr]_+}{\bar n(\bar n -1)(\bar n -2)},
    \quad
    \Sigma_\bullet = \sum \bigl( S_1^3 - 3 S_1 S_2 + 2 S_3 \bigr),
\end{equation}
where $[\,\cdot\,]_+$ clamps at zero against floating-point drift and $\Sigma_c$
uses the column sums, $\Sigma_v$ the row sums. Driving both terms down pushes the
Tanner girth to $8$ or above. \\
The fourth term pushes down the second-largest eigenvalue of the two node projections, each normalized by its diagonal part \citep{910591}
$D_c = \mathrm{diag}(A_c)$, $D_v = \mathrm{diag}(A_v)$:
\begin{equation}
    E_{\mathrm{spectral}} \;=\;
    \mu_2\!\left(D_c^{-1/2} A_c D_c^{-1/2}\right)
    \;+\;
    \mu_2\!\left(D_v^{-1/2} A_v D_v^{-1/2}\right).
\end{equation}
We use this as a heuristic for the code's distance, since it assumes a regular binary graph, whereas during sampling $\mathcal{M}$ is continuous. \\
The fifth term prevents a rank collapse.  We use the log-determinant over the singular values $s_i$:
\begin{equation}
    E_{\mathrm{rank}} \;=\;
    -\frac{1}{\min(\bar m, \bar n)} \sum_{i} \log s_i(\mathcal{M}),
\end{equation}
which diverges as any singular value approaches zero, penalizing directions in which rank is lost.
\paragraph{3. Hardware}
The sixth term aims to keep the check weights light. We
use a soft maximum of the row and column sums of $\mathcal{M}$:
\begin{equation}
    E_{\mathrm{depth}} \;=\;
    \frac{1}{\beta}\log \sum_{i} e^{\beta r_i}
    \;+\;
    \frac{1}{\beta}\log \sum_{j} e^{\beta c_j},
    \qquad
    r_i = \sum_j \mathcal{M}_{ij}, \quad c_j = \sum_i \mathcal{M}_{ij},
\end{equation}
The gradient is distributed across the rows in proportion to the exponential of their sums, so the heaviest rows dominate while every row retains a nonzero contribution. Increasing $\beta$ sharpens the term toward the true maximum. The last two terms act directly on $\widehat{M}_0$ rather than its unfolding. The seventh term is exactly the degree term used in training \ref{eq:train-loss}, holding the
support degrees at their targets during sampling.
The eighth is the plain $L_1$ norm of the base matrix, the expected number of ones, to maintain a general sparse structure for the final output: $E_{L_1} \;=\; \sum_{i,j,k} \widehat{M}_{0,ijk},$
\paragraph{The combined energy}
Finally, we obtain the combined energy:
\begin{equation}
\begin{aligned}
    E \;=\;\;
      &\lambda_{\mathrm{LER}}\, E_{\mathrm{LER}}
    + \lambda_{\mathrm{girth}_4}\, E_{\mathrm{girth}_4}
    + \lambda_{\mathrm{girth}_6}\, E_{\mathrm{girth}_6}
    + \lambda_{\mathrm{spectral}}\, E_{\mathrm{spectral}} \\
    +\;&\lambda_{\mathrm{rank}}\, E_{\mathrm{rank}}
    + \lambda_{\mathrm{depth}}\, E_{\mathrm{depth}}
    + \lambda_{\deg}\, E_{\deg}
    + \lambda_{L_1}\, E_{L_1},
\end{aligned}
\label{eq:energy-full}
\end{equation}
whose gradient is what \eqref{eq:guided} follows. We state the weights used for every reported run. After the reverse loop finishes, one further gradient step is taken directly on the continuous tensor to refine uncertain final probabilities. The tensor is then clipped to $[0,1]$ and rounded at $\tfrac12$. This last step is scaled separately from $g$, and is specified per shape.
\subsection{Screening and exact distance certification}
\label{sec:milp}
Generation returns batches of candidate base matrices. Each generated sample is expanded into the corresponding parity-check matrices $H_X$ and $H_Z$, which are first verified to satisfy the commutation condition
$H_X H_Z^{\mathsf{T}} = 0$. For each valid code, compute its block length $n$ and number of logical qubits
$k = n - \operatorname{rank} H_X - \operatorname{rank} H_Z$.
During evaluation, the distance of each generated code is estimated using the open-source upper-bound tool of \citet{Pryadko_2022}. The generated codes are subsequently ranked according to their $n$, $k$, distance upper bound and maximum check weight, where the latter is defined as the heaviest row weight across $H_X$ and $H_Z$. As exact quantum-code distance computation is NP-hard in general \citep{Kapshikar_2023}, we use MILP to compute the exact distance of the final candidates, using the implementation of \citet{cruzbenito2026evolutionarydiscoverybivariatebicycle}.

\section{Results}
\label{sec:results}

\begin{table}[t]
  \centering
  \caption{The strongest codes generated by the model. $k/n$ is the encoding
  rate and $w_{\mathrm{chk}}$ the maximum check weight, the largest row weight
  over both $H_X$ and $H_Z$.}
  \begin{tabular}{clcc@{\hspace{2.2em}}clcc}
    \hline
    \# & $[[n, k, d]]$ & $k/n$ & $w_{\mathrm{chk}}$ &
    \# & $[[n, k, d]]$ & $k/n$ & $w_{\mathrm{chk}}$ \\
    \hline
    \noalign{\vskip 0.9ex}
    1  & $[[260, 20, 6]]$                 & $0.077$ & $5$ & 17 & $[[156, 20, 8]]$       & $0.128$ & $8$ \\
    2  & $[[260, 20, 8]]$                 & $0.077$ & $6$ & 18 & $\textbf{[[247, 19, 14]]}$       & $0.077$ & $8$ \\
    3  & $[[260, 20, 11]]$                & $0.077$ & $7$ & 19 & $[[325, 45, 8]]$                 & $0.138$ & $9$ \\
    4  & $[[182, 14, 11]]$                & $0.077$ & $7$ & 20 & $[[400, 80, 6]]$                 & $0.200$ & $9$ \\
    5  & $[[247, 19, 11]]$                & $0.077$ & $7$ & 21 & $[[312, 24, 10]]$                & $0.077$ & $9$ \\
    6  & $\textbf{[[169, 13, 11]]}$       & $0.077$ & $7$ & 22 & $[[169, 13, 12]]$                & $0.077$ & $9$ \\
    7  & $[[592, 32, 13]]$                & $0.054$ & $7$ & 23 & $[[493, 153, 6]]$                & $0.310$ & $9$ \\
    8  & $[[65, 5, 7]]$                   & $0.077$ & $7$ & 24 & $\textbf{[[340, 68, 9]]}$                 & $0.200$ & $9$ \\
    9  & $[[117, 9, 9]]$                  & $0.077$ & $7$ & 25 & $[[160, 32, 6]]$                 & $0.200$ & $9$ \\
    10 & $[[91, 13, 5]]$        & $0.143$ & $7$ & 26 & $[[195, 27, 6]]$                 & $0.138$ & $10$ \\
    11 & $[[592, 32, 14]]$                & $0.054$ & $8$ & 27 & $[[480, 96, 7]]$                 & $0.200$ & $10$ \\
    12 & $[[91, 7, 8]]$                   & $0.077$ & $8$ & 28 & $[[267, 27, 7]]$                 & $0.101$ & $10$ \\
    13 & $\textbf{[[260, 52, 7]]}$        & $0.200$ & $8$ & 29 & $\textbf{[[730, 180, 7]]}$       & $0.247$ & $11$ \\
    14 & $\textbf{[[260, 20, 12]]}$       & $0.077$ & $8$ & 30 & $[[260, 20, 14]]$                & $0.077$ & $11$ \\
    15 & $[[260, 24, 10]]$      & $0.092$ & $8$ & 31 & $[[493, 153, 7]]$                & $0.310$ & $11$ \\
    16 & $[[143, 15, 9]]$       & $0.105$ & $8$ & 32 & $[[493, 153, 8]]$                & $0.310$ & $12$ \\
    \hline
  \end{tabular}
  \label{tab:best-codes}
\end{table}
\subsection{Experimental setup}
\label{sec:setup}

Our experimental evaluation follows the pipeline introduced in Section~\ref{sec: method}. We employ the network described in Section~\ref{sec:model} as the denoiser in the diffusion process. The architecture consists of $L$ layers, $h$ attention heads, and an embedding dimension $d$. While the exact configuration varies per setting, the parameter count is approximately $10.7$M, depending slightly on $\ell$. Training samples are generated on the fly. Due to the lack of relevant data and to provide a more structured training distribution than uniformly random matrices, each sample is constructed in two steps. First, a binary $m\times n$ support is generated using a PEG procedure \citep{1377521}. Second, each nonzero entry in the support is assigned a random polynomial with weight $w\sim\mathrm{Uniform}\{1,\dots,w_{\max}\}$, with $w$ positions selected uniformly. We treat $w_{\max}$ as a hyperparameter, with the goal of providing the model with a basic sparse structural prior. The forward corruption process spans $T=100$ steps under a cosine schedule, with $\beta_t$ ranging from $10^{-4}$ to $2\times10^{-2}$. We train the model for $750$ epochs using AdamW \citep{loshchilov2017decoupled} with a batch size of $16$. In each epoch, $4096$ base matrices are generated using the PEG sampler. The peak learning rate is $10^{-4}$ with a weight decay of $10^{-4}$. The learning rate undergoes a linear warmup over the first $20$ epochs before annealing to zero via cosine decay. Gradients are clipped to a maximum norm of $1.0$, and the degree term in the loss function of equation \ref{eq:train-loss} is weighted by $\lambda_{\deg}=3.0$. Training is performed on a single NVIDIA L40 GPU, with each epoch taking approximately $5$ seconds. During inference, each run generates a batch of $64$ samples using the guided reverse process defined in \eqref{eq:guided}. Each run is initialized with a specific base geometry $(m,n,\ell)$, a target column weight $d_v$, and a maximum polynomial weight $w_{\max}$. The guidance scale and the hyperparameters of all energy terms are tuned individually for each run. Complete configurations for all generated results are provided in Appendix~\ref{app: Full Codes}.

\begin{figure}[t]
  \centering
  \includegraphics[width=0.85\textwidth]{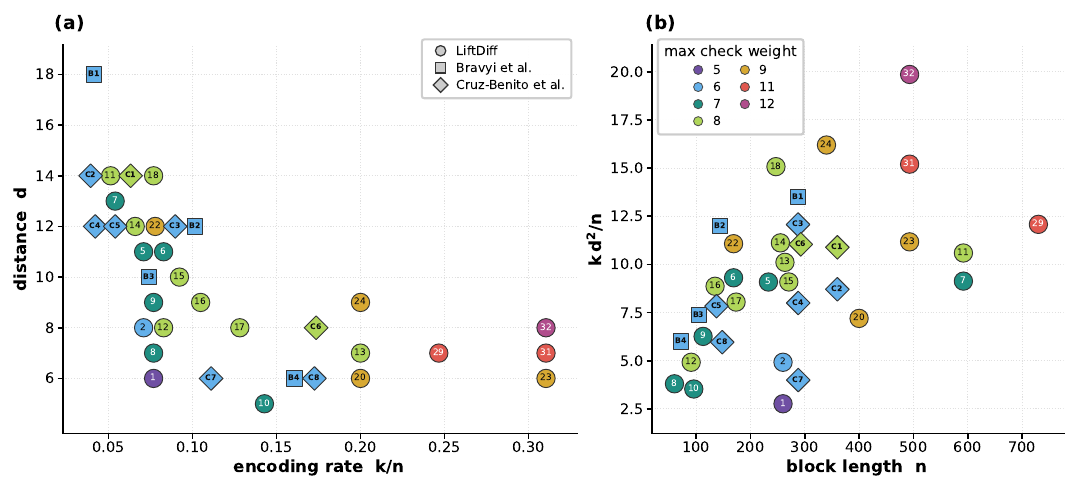} 
  \caption{Selected codes from Table~\ref{tab:best-codes} against the baselines.
  Each marker is one code, colored by its $w_{chk}$. Circles are
  codes from this work, tagged with their row number in
  Table~\ref{tab:best-codes}. Squares are BB codes of
  \citet{Bravyi_2024} and diamonds the CSS codes of \citet{cruzbenito2026evolutionarydiscoverybivariatebicycle}.
  \textbf{(a)} Encoding rate $k/n$ against distance $d$.
  \textbf{(b)} Block length $n$ against $k d^{2}/n$.
  Codes sharing a point exactly are offset by a small fraction of the axis.
  Baseline key: B1 $[[288,12,18]]$, B2 $[[144,12,12]]$, B3 $[[108,8,10]]$,
  B4 $[[72,12,6]]$;
  C1 $[[360,20,\le 14]]$, C2 $[[360,16,\le 14]]$, C3 $[[288,24,12]]$,
  C4 $[[288,16,12]]$, C5 $[[144,8,12]]$, C6 $[[288,50,8]]$, C7 $[[288,32,6]]$,
  C8 $[[144,24,6]]$.}
  \label{fig:codes}
\end{figure}

\subsection{Generated codes}
\label{sec:codes}
We gathered 32 codes produced by LiftDiff, detailed in the accompanying Table~\ref{tab:best-codes}. The discovered codes cover a broad range of $[[n, k, d]]$ parameters, with block lengths ($n$) ranging from 65 to 730, and distances ($d$) between 5 and 14. Our construction achieves high encoding rates ($k/n$), reaching up to 0.310 while keeping the maximum check weight ($w_{\text{chk}}$) within a range of 5 to 12. Several codes demonstrate a favorable balance between encoding rate, distance, and hardware-friendly check weights (for example 6, 13--16 and 18). To evaluate the performance of our construction, we benchmarked the new codes against the BB codes \citep{Bravyi_2024} and recent work \citep{cruzbenito2026evolutionarydiscoverybivariatebicycle} in Figure~\ref{fig:codes}. Panel (a) illustrates the trade-off between $d$ and $k/n$. Our codes achieve high encoding rates, extending beyond 0.30. Code 18 occupies a favorable position, achieving high rate, $d=14$ and $w_{chk} = 8$, compared to C1 and C2 which achieve lower rate and the same distance. Panel (b) evaluates the overall code efficiency $kd^2/n$ as a function of the block length $n$. The results demonstrate that our method achieves high code efficiency across a range of block lengths. Notably, our codes reach $kd^2/n$ values approaching 17 (code 24), significantly outperforming C1 and C2 while increasing $w_{chk}$ by only one. Additionally, code 18 achieves the highest $kd^2/n$ with $w_{chk} \le 8$, at a modest block length within the range of the baseline codes.

\subsection{Logical error rate}
\label{sec:ler}
\begin{figure}[t]
  \centering
  \begin{subfigure}{0.28\textwidth}
    \centering
    \includegraphics[width=\linewidth]{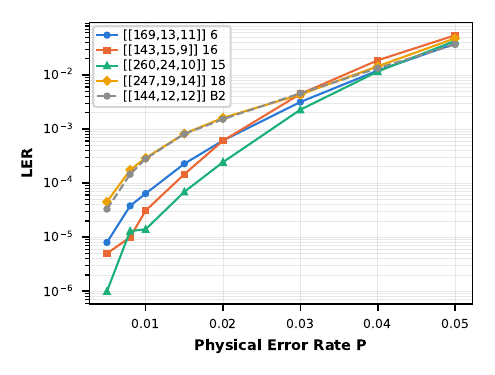}
    \caption{Code capacity LER}
    \label{fig:LER_cap}
  \end{subfigure}\hfill
    \begin{subfigure}{0.4\textwidth}
    \centering
    \includegraphics[width=\linewidth]{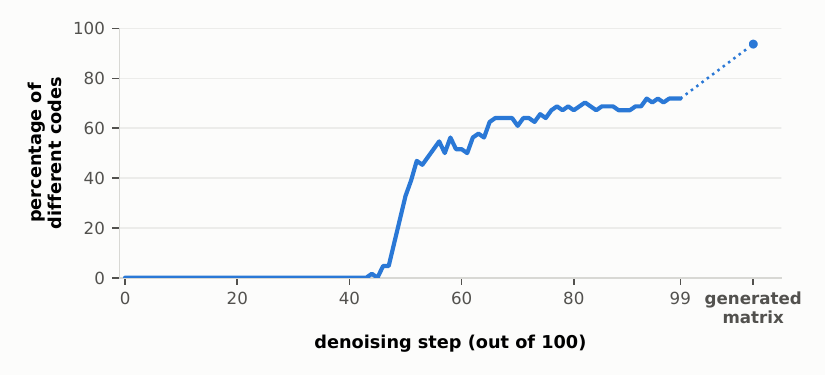}
    \caption{Code divergence over steps}
    \label{fig:div}
  \end{subfigure}\hfill
  \begin{subfigure}{0.28\textwidth}
    \centering
    \includegraphics[width=\linewidth]{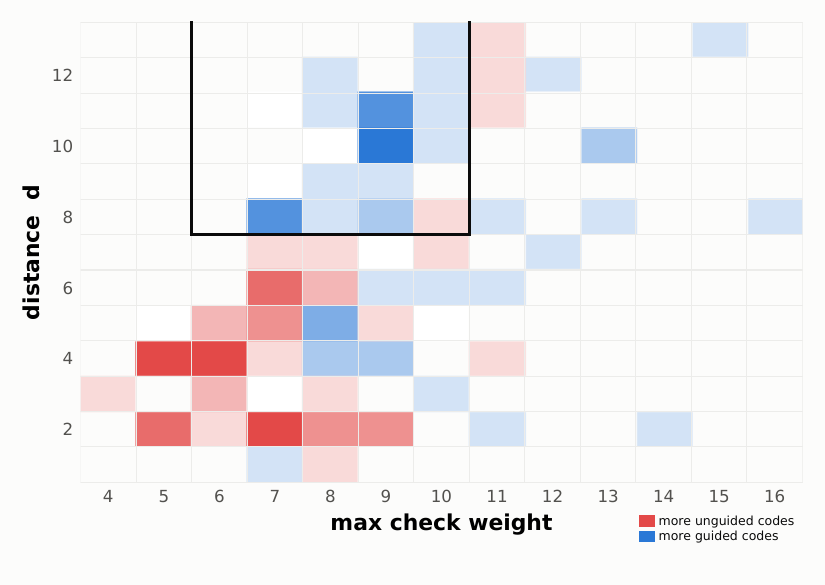}
    \caption{Guided vs unguided codes}
    \label{fig:2D_histogram}
  \end{subfigure}
  \caption{\textbf{(a)} Code-capacity block LER vs. physical error rate $p$ for four LiftDiff codes and the [[144,12,12]] BB code. Each label's index identifies the code in Table~\ref{tab:best-codes}. \textbf{(b)} Share of codes that differ from their unguided counterpart per denoising step. \textbf{(c)} Weight-distance cells where guidance produces more codes (blue) or fewer (red) than unguided generation. Distances are upper bounds.}
  \label{fig:ler}
\end{figure}

To evaluate the error-correction performance of the codes generated by LiftDiff, we measure the code-capacity logical error rate (LER) as a
function of the physical error rate $p$ as illustrated in Figure~\ref{fig:LER_cap}. We consider independent $X$ errors occurring on each qubit with probability $p$, and decode the resulting $Z$-syndrome using BP-OSD with product-sum BP
for 20 iterations, followed by OSD with CS order 7, following the
simulation procedure of~\citet{cruzbenito2026evolutionarydiscoverybivariatebicycle}. For each value of $p$, we generate $10^6$ independent error samples. Figure~\ref{fig:LER_cap} shows the resulting LER curves for four codes generated by LiftDiff, together with the $\left[\left[144,12,12\right]\right]$ code for comparison. The $[[260,24,10]]$ and $[[169,13,11]]$ codes stay below the $[[144,12,12]]$ curve over almost the entire range, up to $p=0.04$, and only meet the baseline at $p=0.05$. The $[[143,15,9]]$ code is below it for $p \le 0.02$, matches it at $p=0.03$ and rises above it beyond that. The gap is widest at low physical error rates: at $p=0.01$ the $[[260,24,10]]$ code reaches an LER about twenty times lower than the $[[144,12,12]]$ code, while $[[143,15,9]]$ and $[[169,13,11]]$ are lower by factors of about nine and four. The $[[247,19,14]]$ code behaves differently, and follows the $[[144,12,12]]$ curve closely across the whole range. The LER reported here is a block error rate, a sample is counted as a failure if any of the $k$ logical qubits is flipped after decoding, rather than averaging over the individual logical qubits. This makes the result of the $[[260,24,10]]$ code notable, since it reaches the lowest curve while encoding twice as many logical qubits as the $[[144,12,12]]$ code. Appendix \ref{app:circuit} provides circuit-level evaluation.

\section{Model Analysis}
To further evaluate the proposed method, we conducted additional experiments. In Figure~\ref{fig:div}, we present two runs initialized with the same seed, one without the guidance mechanism and the other with guidance. The curve shows the percentage of differing matrices between the two batches throughout the denoising process. Since the two runs are otherwise identical, each matrix in one batch can be paired with the corresponding matrix in the other batch, allowing us to directly measure the effect of guidance on each sample. As the denoising process progresses, the percentage of differing matrices increases, such that by the end of the process, there are almost no identical matrices between the two batches. In Figure~\ref{fig:2D_histogram}, we examine how this divergence is reflected in the distance and maximum check weight of the resulting codes. The target region corresponds to high distance and a maximum check weight between 6 and 10, as indicated by the marked boundaries. This region is predominantly occupied by codes generated with the guidance mechanism, aggregated across multiple random seeds. In contrast, majority of codes generated by the model without guidance are concentrated at lower maximum check weights, but also exhibit lower distances. This separation indicates that, while the unguided model tends to generate codes with lower check weights, the guidance mechanism steers the generation process toward regions of the search space that are more relevant to our target criteria, allowing us to reach codes with higher distances while maintaining a maximum check weight in the desired range.

\section{Conclusion}
In this work, we introduced LiftDiff, the first discrete diffusion model designed for generating QECC, specifically, LP codes. Our framework demonstrates that generative models provide a highly effective strategy for navigating the vast search space of QECC. The experimental results yield robust codes characterized by competitive encoding rates, code distances, and low LER.

\bibliography{references} 
\bibliographystyle{iclr2027_conference}

\appendix

\section{Belief-propagation configuration}
\label{app:bp}

$E_{\mathrm{LER}}$ uses sum--product BP, unrolled for $20$
iterations and run once per error type ($H_Z$ for $e_X$, $H_X$ for $e_Z$). All
messages are recomputed in parallel at each iteration, and the loop always runs
to its full depth rather than stopping once the syndrome is matched.
Variable-to-check messages are damped: each is $0.8$ of its newly computed value
plus $0.2$ of its previous one. Every edge is weighted by the corresponding
entry of $H_X$ or $H_Z$, so an entry of $1$ gives the ordinary update and an
entry of $0$ removes the edge, which is what lets the decoder run on the relaxed
check matrices. The physical error rate is $p = 0.05$, so the channel
log-likelihood ratio is $3.37$ on every variable node, and each evaluation of
$E_{\mathrm{LER}}$ averages $64$ sampled error patterns. The arguments of
$\tanh$ and $\operatorname{artanh}$ are clamped to $\pm 0.9999$ and the
posterior to $[10^{-6}, 1-10^{-6}]$. All of these are fixed for every run.

\section{Circuit-level evaluation}
\label{app:circuit}

\begin{figure}[t]
  \centering
  \includegraphics[width=0.4\linewidth]{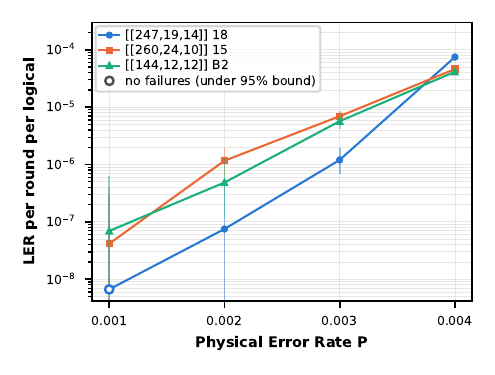}
  \caption{Circuit-level LER per logical qubit per round vs. physical
  error rate $p$ for two codes produced by LiftDiff and the $[[144,12,12]]$ BB
  code. Number of rounds
  equal to each code's distance. The open marker denotes a point with no observed failures, showing the $95\%$ upper bound.
  The index beside each label identifies the code in Table~\ref{tab:best-codes}.}
  \label{fig:circuit-ler}
\end{figure}

To evaluate the performance of the generated codes under circuit-level noise, we
measure the LER per logical qubit per round as a
function of the physical error rate $p$. The experiment covers the
$[[247,19,14]]$ and $[[260,24,10]]$ codes of Table~\ref{tab:best-codes},
together with the $[[144,12,12]]$ code as a baseline. Every code is compiled into a syndrome-extraction circuit and its detector error
model under the same protocol \citep{perlin2023qldpc}, to provide an equal comparison. We repeat the syndrome measurement for a number of rounds equal to the code distance, $14$, $10$ and $12$ respectively, and measure once in the $X$ basis and once in the $Z$ basis. We decode with the relay BP decoder \citep{müller2025improvedbeliefpropagationsufficient} with a stop after the first converged solution, under circuit-level depolarizing noise of strength $p$ for $p \in \{0.001, 0.002, 0.003, 0.004\}$. Each
point draws up to $10^5$ shots per basis and stops once it has recorded $200$ failures.
Figure~\ref{fig:circuit-ler} shows the resulting curves. The $[[247,19,14]]$
code stays below the $[[144,12,12]]$ curve from $p=0.001$ to $p=0.003$. It
records no failures at $p=0.001$,  At $p=0.004$ the baseline takes a small
margin. The $[[260,24,10]]$ code follows the
$[[144,12,12]]$ curve closely over the whole range and stays within its confidence intervals at every point.

\section{Detailed Codes Configuration}
\label{app: Full Codes}
\ifdefined\dcruntable\else\newsavebox{\dcruntable}\fi
\begin{table}[t]
\caption{Every code of Table~\ref{tab:best-codes} together with the
  settings of the run that produced it.
  $m \times n \times \ell$ is the base geometry, $d_v$ the target column weight
  and $w_{\max}$ the largest polynomial weight the training data was built with.
  $g$ is the guidance scale of~\eqref{eq:guided} and $g_{\mathrm{fin}}$ the
  scale of the single gradient step taken on the continuous tensor before
  binarization. $=g$ marks configuration where the same scale was used in both.  The remaining columns are the energy weights of~\eqref{eq:energy-full}. Every other setting was the same in all 
  experiments: 100 diffusion steps on a cosine $\beta$ schedule from
  $10^{-4}$ to $2 \times 10^{-2}$, a 256-wide transformer with
  4 heads and 6 layers, except where a row is marked. 750 epochs of
  4096 base matrices produced by PEG, in batches of 16, optimized by AdamW at learning rate $10^{-4}$ with
  weight decay $10^{-4}$, 20 warm-up epochs, cosine annealing and
  gradient clipping at 1, a training-time degree weight of 3 with margin 0, and 64 samples drawn.}
  \label{tab:run-settings}
  \centering
  \setlength{\tabcolsep}{2.5pt}
  \sbox{\dcruntable}{
  \begin{tabular}{r l @{\hspace{7pt}} c c c @{\hspace{7pt}} c c @{\hspace{7pt}} c c c c c c c c}
    \hline
    \# & $[[n , k, d]]$ & $m\times n\times\ell$ & $d_v$ & $w_{\max}$ & $g$ & \rotatebox{90}{$g_{\mathrm{fin}}$} & \rotatebox{90}{$\lambda_{\mathrm{LER}}$} & \rotatebox{90}{$\lambda_{\mathrm{girth}_4}$} & \rotatebox{90}{$\lambda_{\mathrm{girth}_6}$} & \rotatebox{90}{$\lambda_{\mathrm{spectral}}$} & \rotatebox{90}{$\lambda_{\mathrm{rank}}$} & \rotatebox{90}{$\lambda_{\mathrm{depth}}$} & \rotatebox{90}{$\lambda_{\deg}$} & \rotatebox{90}{$\lambda_{L_1}$} \\
    \hline
    \noalign{\vskip 0.6ex}
    1 & $[[260, 20, 6]]$ & $4\times 6\times 5$ & 2 & 1 & 2 & $=g$ & 15 & 0.15 & 0.015 & 0.4 & 0.2 & 0.1 & 0.25 & 0.025 \\
    2 & $[[260, 20, 8]]$ & $4\times 6\times 5$ & 2 & 1 & 2 & $=g$ & 1 & 0.15 & 0.015 & 0.25 & 0.5 & 0.1 & 0.25 & 0.02 \\
    3$^{\mathrm{ab}}$ & $[[260, 20, 11]]$ & $2\times 3\times 20$ & 2 & 3 & 2 & $=g$ & 10 & 0.15 & 0.01 & 0.5 & 0.5 & 0.1 & 0.25 & 0.01 \\
    4 & $[[182, 14, 11]]$ & $2\times 3\times 14$ & 2 & 2 & 1 & 0.5 & 10 & 0.15 & 0.01 & 0.5 & 0.5 & 0.1 & 0.25 & 0.02 \\
    5 & $[[247, 19, 11]]$ & $2\times 3\times 19$ & 2 & 2 & 2 & 0.5 & 10 & 0.15 & 0.015 & 0.6 & 0.5 & 0.1 & 0.3 & 0.02 \\
    6 & $[[169, 13, 11]]$ & $2\times 3\times 13$ & 2 & 3 & 1 & 0.5 & 10 & 0.15 & 0.01 & 0.5 & 0.5 & 0.1 & 0.25 & 0.02 \\
    7 & $[[592, 32, 13]]$ & $5\times 7\times 8$ & 3 & 1 & 2 & $=g$ & 15 & 0.15 & 0.015 & 0.5 & 0.3 & 0.1 & 0.25 & 0.01 \\
    8 & $[[65, 5, 7]]$ & $2\times 3\times 5$ & 2 & 1 & 3 & $=g$ & 10 & 0.15 & 0.01 & 0.5 & 0.5 & 0.1 & 0.25 & 0.03 \\
    9 & $[[117, 9, 9]]$ & $2\times 3\times 9$ & 2 & 2 & 2 & 0.5 & 10 & 0.15 & 0.01 & 0.5 & 0.5 & 0.1 & 0.25 & 0.02 \\
    10 & $[[91, 13, 5]]$ & $2\times 3\times 7$ & 2 & 2 & 1 & 0.5 & 10 & 0.15 & 0.01 & 0.5 & 0.5 & 0.1 & 0.25 & 0.04 \\
    11 & $[[592, 32, 14]]$ & $5\times 7\times 8$ & 3 & 1 & 2 & $=g$ & 15 & 0.15 & 0.015 & 0.5 & 0.3 & 0.1 & 0.25 & 0.025 \\
    12 & $[[91, 7, 8]]$ & $2\times 3\times 7$ & 2 & 2 & 1 & 0.5 & 10 & 0.15 & 0.01 & 0.5 & 0.5 & 0.1 & 0.25 & 0.04 \\
    13 & $[[260, 52, 7]]$ & $2\times 4\times 13$ & 2 & 2 & 2 & 0.5 & 10 & 0.15 & 0.01 & 0.5 & 0.5 & 0.1 & 0.25 & 0.03 \\
    14$^{\mathrm{ab}}$ & $[[260, 20, 12]]$ & $2\times 3\times 20$ & 2 & 2 & 1 & $=g$ & 10 & 0.15 & 0.01 & 0.5 & 0.5 & 0.1 & 0.25 & 0.01 \\
    15 & $[[260, 24, 10]]$ & $2\times 3\times 20$ & 2 & 3 & 1 & 0.5 & 10 & 0.15 & 0.01 & 0.7 & 0.5 & 0.1 & 0.25 & 0.03 \\
    16$^{\mathrm{c}}$ & $[[143, 15, 9]]$ & $2\times 3\times 11$ & 2 & 1 & 1 & 0.5 & 10 & 0.15 & 0.015 & 0.5 & 0.2 & 0.1 & 0.2 & 0.01 \\
    17 & $[[156, 20, 8]]$ & $2\times 3\times 12$ & 2 & 2 & 1 & 0.5 & 10 & 0.15 & 0.015 & 0.5 & 0.2 & 0.1 & 0.2 & 0.04 \\
    18 & $[[247, 19, 14]]$ & $2\times 3\times 19$ & 2 & 2 & 5 & 0.5 & 10 & 0.15 & 0.015 & 0.55 & 0.5 & 0.1 & 0.3 & 0.03 \\
    19 & $[[325, 45, 8]]$ & $4\times 7\times 5$ & 3 & 1 & 2 & $=g$ & 10 & 0.15 & 0.015 & 0.5 & 0.5 & 0.1 & 0.25 & 0.02 \\
    20 & $[[400, 80, 6]]$ & $4\times 8\times 5$ & 3 & 1 & 2 & $=g$ & 20 & 0.15 & 0.015 & 0.6 & 0.3 & 0.11 & 0.25 & 0.025 \\
    21 & $[[312, 24, 10]]$ & $4\times 6\times 6$ & 3 & 1 & 2 & $=g$ & 10 & 0.15 & 0.015 & 0.5 & 0.5 & 0.1 & 0.25 & 0.02 \\
    22 & $[[169, 13, 12]]$ & $2\times 3\times 13$ & 2 & 3 & 1 & 0.5 & 10 & 0.15 & 0.01 & 0.5 & 0.5 & 0.1 & 0.25 & 0.02 \\
    23 & $[[493, 153, 6]]$ & $2\times 5\times 17$ & 2 & 2 & 1 & 0.5 & 10 & 0.15 & 0.01 & 0.5 & 0.5 & 0.1 & 0.25 & 0.02 \\
    24 & $[[340, 68, 9]]$ & $2\times 4\times 17$ & 2 & 2 & 2 & 0.5 & 10 & 0.15 & 0.01 & 0.5 & 0.5 & 0.1 & 0.25 & 0.03 \\
    25 & $[[160, 32, 6]]$ & $2\times 4\times 8$ & 2 & 2 & 2 & 0.5 & 10 & 0.15 & 0.01 & 0.5 & 0.5 & 0.1 & 0.25 & 0.02 \\
    26 & $[[195, 27, 6]]$ & $4\times 7\times 3$ & 3 & 1 & 2 & $=g$ & 1 & 0.1 & 0 & 0.5 & 0.5 & 0.1 & 0.25 & 0 \\
    27 & $[[480, 96, 7]]$ & $4\times 8\times 6$ & 3 & 1 & 2 & $=g$ & 1 & 0.15 & 0.015 & 0.25 & 0.5 & 0.1 & 0.25 & 0.02 \\
    28 & $[[267, 27, 7]]$ & $5\times 8\times 3$ & 3 & 1 & 2 & $=g$ & 1 & 0.15 & 0.015 & 0.25 & 0.5 & 0.1 & 0.25 & 0.02 \\
    29 & $[[730, 180, 7]]$ & $5\times 11\times 5$ & 3 & 1 & 2 & $=g$ & 15 & 0.15 & 0.015 & 0.5 & 0.3 & 0.1 & 0.25 & 0.02 \\
    30 & $[[260, 20, 14]]$ & $2\times 3\times 20$ & 2 & 2 & 2 & 0.5 & 10 & 0.15 & 0.015 & 0.6 & 0.5 & 0.1 & 0.3 & 0.04 \\
    31 & $[[493, 153, 7]]$ & $2\times 5\times 17$ & 2 & 2 & 2 & 0.5 & 10 & 0.15 & 0.01 & 0.7 & 0.5 & 0.1 & 0.25 & 0.04 \\
    32 & $[[493, 153, 8]]$ & $2\times 5\times 17$ & 2 & 2 & 1 & 0.5 & 10 & 0.15 & 0.01 & 0.7 & 0.5 & 0.1 & 0.25 & 0.04 \\
    \hline
  \end{tabular}}%
  \ifdim\wd\dcruntable>\linewidth
  \else
    \usebox{\dcruntable}%
  \fi
  \vspace{0.8ex}
  \begin{minipage}{0.97\linewidth}\raggedright\footnotesize $^{\mathrm{a}}$~1 attention head. $^{\mathrm{b}}$~3 layers. $^{\mathrm{c}}$~learning rate $9 \times 10^{-5}$.
  \end{minipage}
\end{table}

\end{document}